\documentclass[aps,prd,twocolumn,superscriptaddress,showpacs,preprintnumbers,amsmath,amssymb]{revtex4}
\usepackage{graphicx}
\usepackage{dcolumn}
\usepackage{amsmath}
\usepackage{bm}

\begin{document}


\title{Observation of $B^{\pm} \to \chi _{c1} \pi ^{\pm}$ and Search for Direct $CP$ Violation}

\affiliation{Budker Institute of Nuclear Physics, Novosibirsk}
\affiliation{Chiba University, Chiba}
\affiliation{Chonnam National University, Kwangju}
\affiliation{University of Cincinnati, Cincinnati, Ohio 45221}
\affiliation{University of Hawaii, Honolulu, Hawaii 96822}
\affiliation{High Energy Accelerator Research Organization (KEK), Tsukuba}
\affiliation{University of Illinois at Urbana-Champaign, Urbana, Illinois 61801}
\affiliation{Institute of High Energy Physics, Chinese Academy of Sciences, Beijing}
\affiliation{Institute of High Energy Physics, Vienna}
\affiliation{Institute of High Energy Physics, Protvino}
\affiliation{Institute for Theoretical and Experimental Physics, Moscow}
\affiliation{J. Stefan Institute, Ljubljana}
\affiliation{Kanagawa University, Yokohama}
\affiliation{Korea University, Seoul}
\affiliation{Kyungpook National University, Taegu}
\affiliation{Swiss Federal Institute of Technology of Lausanne, EPFL, Lausanne}
\affiliation{University of Ljubljana, Ljubljana}
\affiliation{University of Maribor, Maribor}
\affiliation{University of Melbourne, Victoria}
\affiliation{Nagoya University, Nagoya}
\affiliation{Nara Women's University, Nara}
\affiliation{National Central University, Chung-li}
\affiliation{National United University, Miao Li}
\affiliation{Department of Physics, National Taiwan University, Taipei}
\affiliation{H. Niewodniczanski Institute of Nuclear Physics, Krakow}
\affiliation{Nippon Dental University, Niigata}
\affiliation{Niigata University, Niigata}
\affiliation{Osaka City University, Osaka}
\affiliation{Osaka University, Osaka}
\affiliation{Panjab University, Chandigarh}
\affiliation{Peking University, Beijing}
\affiliation{University of Pittsburgh, Pittsburgh, Pennsylvania 15260}
\affiliation{Princeton University, Princeton, New Jersey 08544}
\affiliation{RIKEN BNL Research Center, Upton, New York 11973}
\affiliation{Saga University, Saga}
\affiliation{University of Science and Technology of China, Hefei}
\affiliation{Sungkyunkwan University, Suwon}
\affiliation{University of Sydney, Sydney NSW}
\affiliation{Tata Institute of Fundamental Research, Bombay}
\affiliation{Toho University, Funabashi}
\affiliation{Tohoku Gakuin University, Tagajo}
\affiliation{Tohoku University, Sendai}
\affiliation{Department of Physics, University of Tokyo, Tokyo}
\affiliation{Tokyo Institute of Technology, Tokyo}
\affiliation{Tokyo Metropolitan University, Tokyo}
\affiliation{Tokyo University of Agriculture and Technology, Tokyo}
\affiliation{Virginia Polytechnic Institute and State University, Blacksburg, Virginia 24061}
\affiliation{Yonsei University, Seoul}
   \author{R.~Kumar}\affiliation{Panjab University, Chandigarh} 
   \author{J.~B.~Singh}\affiliation{Panjab University, Chandigarh} 
     \author{K.~Abe}\affiliation{High Energy Accelerator Research Organization (KEK), Tsukuba} 
   \author{K.~Abe}\affiliation{Tohoku Gakuin University, Tagajo} 
   \author{H.~Aihara}\affiliation{Department of Physics, University of Tokyo, Tokyo} 
   \author{D.~Anipko}\affiliation{Budker Institute of Nuclear Physics, Novosibirsk} 
   \author{T.~Aushev}\affiliation{Institute for Theoretical and Experimental Physics, Moscow} 
   \author{T.~Aziz}\affiliation{Tata Institute of Fundamental Research, Bombay} 
   \author{A.~M.~Bakich}\affiliation{University of Sydney, Sydney NSW} 
   \author{V.~Balagura}\affiliation{Institute for Theoretical and Experimental Physics, Moscow} 
% ...

\collaboration{The Belle Collaboration}
\noaffiliation

\begin{abstract}
We report the first observation of $B^{\pm} \to \chi_{c1} \pi^{\pm}$, a Cabibbo- and color-suppressed decay  in a data sample of $386\times 10^6$~$B\overline B$ events collected at the $\Upsilon(4S)$ resonance with the Belle detector at the KEKB asymmetric-energy $e^+e^-$ collider. We observe $55\pm10$ signal events with a statistical significance of $6.3\sigma$ including systematic uncertainties. The measured branching fraction and charge-asymmetry is $\mathcal {B}( B^{\pm} \to \chi_{c1} \pi^{\pm}) = (2.2\pm 0.4\pm 0.3)\times 10^{-5}$ and $\mathcal{A}_{\pi} = 0.07 \pm 0.18\pm 0.02$, respectively. We also determine the ratio $\mathcal {B}(B^{\pm} \to \chi_{c1} \pi^{\pm})/\mathcal {B}(B^{\pm} \to \chi_{c1} K^{\pm}) = (4.3\pm 0.8\pm 0.3) \%$.
\end{abstract}

\maketitle

 Decays of $B$-mesons to two-body final states including charmonium are expected to occur predominantly via the color-suppressed spectator diagram as shown in Fig. 1. The branching fraction for  the $B^-\to \chi_{c1} K^{-}$~\cite{foot} decay mode is well measured by Belle and BaBar~\cite{bellechic,babar}. To produce this final state, the vector current ($W^-$) couples to a $\bar {c}s$ pair; the $s$-quark and spectator anti-quark hadronize into a kaon. If this theoretical description is correct, a corresponding Cabibbo-suppressed decay mode should exist, where the vector current
couples to a $\bar {c}d$ pair. The $d$-quark and spectator anti-quark
hadronize as a pion, which leads to a $B^- \to \chi_{c1} \pi^-$ decay. If the leading-order tree level diagram is the dominant contribution, the factorization picture implies that the branching fraction of $B^- \to \chi_{c1}\pi^-$ decay mode should be $\sim 5\%$ of that of the Cabibbo-allowed $B^- \to \chi_{c1}K^-$ decay mode~\cite{five}. The Standard Model predicts that for $b \to c\bar cs$ decays, the tree and penguin contributions have a small relative weak phase.  Therefore, negligible direct $CP$-violation is expected in $B^{\pm} \to \chi_{c1}K^{\pm}$ decay. In $b \to c\bar cd$ transitions, however, tree and penguin contributions have different phases and direct $CP$-violation may be as large as a few percent ~\cite{babar1,babar2}. 

In this paper, we report the first observation of  $B^- \to \chi_{c1} \pi^-$ decay. A measurement of the ratio of branching fractions $\mathcal {B} (B^-\to \chi _{c1} \pi^-)/ \mathcal {B}(B^-\to \chi _{c1} K^- )$  and a search for direct $CP$-violation in $B^{\pm} \to \chi_{c1} \pi^{\pm}$ decays is also presented. We use a  data sample containing $(386 \pm 5) \times ~ 10^6~B\overline B$ events collected at the $\Upsilon(4S)$ resonance with the  Belle detector~\cite{belle} at the KEKB asymmetric-energy $e^+e^-$ collider~\cite{kekb}.
\begin{figure}
\begin{tabular}{c}
\includegraphics[width=0.4\textwidth]{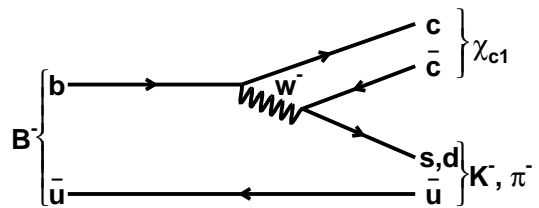}
\end{tabular}
\caption{Leading-order tree level diagram for the decays under study.}
\end{figure}

 The Belle detector is a large solid-angle magnetic spectrometer located at the KEKB $e^+ e^-$ storage rings, which collide 8.0 GeV electrons with 3.5 GeV positrons producing a center-of-mass (CM) energy of 10.58 GeV, the mass of the $\Upsilon (4S)$ resonance.
Closest to the interaction point (IP) is
a silicon vertex detector (SVD),
surrounded by a 50-layer central drift chamber (CDC), 
an array of aerogel Cherenkov counters (ACC),
a barrel-like arrangement of time-of-flight (TOF) scintillation counters, 
 and an electromagnetic calorimeter (ECL) comprised of CsI(T$l$)
 crystals. 
These subdetectors are located inside a superconducting solenoid coil
 that provides a 1.5~T magnetic field. An iron flux-return yoke located outside
 the coil is instrumented to detect $K^0_L$ mesons and to identify muons. The detector is described in detail elsewhere~\cite{belle}. The data set consists of two subsets: the first $152 \times 10^{6}~$ $B$-meson pairs were collected with a 2.0 cm radius beam-pipe and a 3-layer SVD, and the remaining $234 \times 10^{6}~$ $B$-meson pairs with a 1.5 cm radius beam-pipe, a 4-layer SVD and a small-cell inner drift chamber~\cite{yoshi}.

Events with $B$-meson candidates are first selected by applying
general hadronic event selection criteria.  These include a requirement on
charged tracks (at least 
     three of them should originate from an event vertex consistent
with the IP), a requirement on the reconstructed CM
energy ($E^{CM} > 0.2 \sqrt{s}$, where $\sqrt{s}$ is the total CM energy), a
requirement on the longitudinal 
     ($z$-direction) component of the reconstructed CM momentum with
respect to the beam 
     direction ($|p^{CM}_z| < 0.5 \sqrt{s}/c$), and a requirement on
the total ECL energy 
     ($0.1\sqrt{s} < E^{CM}_{ECL} < 0.8\sqrt{s}$) with at least two
energy clusters.
To suppress continuum background, we reject events where the  
ratio of the second to zeroth Fox-Wolfram moments~\cite{Fox78} is greater than
0.5.  To remove charged particle tracks that are poorly
measured or do not come from the interaction region, we require their origin to be within $0.5~\rm cm$ of the IP in the radial direction, and $5~\rm cm$ along the beam direction ($z$-direction).
\begin{figure}
\resizebox{8.8cm}{7cm}{\includegraphics{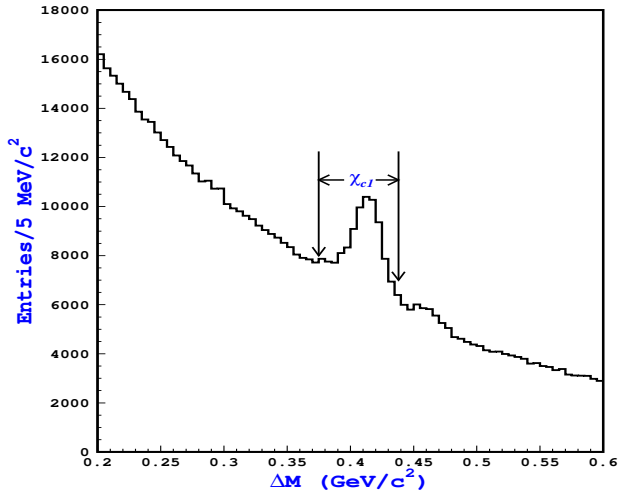}}
\caption{The $\Delta M$ ($M_{\ell ^+ \ell ^-\gamma} - M_{\ell^+\ell^-}$) distribution for the $\chi _{c1}$ candidates. The arrows indicate the selected mass region. The enhancement just above the $\chi_{c1}$ mass region is due to the $\chi_{c2}$.}\end{figure}

We reconstruct the ${\chi}_{c1}$ state via the decay mode
${\chi}_{c1}\to \gamma J/\psi$. We begin by reconstructing $J/\psi \to \ell^+\ell^-$ candidates, where $\ell$ is a muon or electron. For muon tracks, identification is based on track
penetration depth and the hit pattern in the KLM system. 
Electron tracks
are identified by a combination of $dE/dx$ from the CDC, $E/p$
($E$ is the energy deposited in the ECL and $p$ is the momentum measured
by the SVD and the CDC), and shower shape in the ECL. In order to
recover di-electron events in which one or both electrons 
radiate a photon, the four-momenta of all photons within 0.05
radians of the $e^+$ or $e^-$ directions are included in the invariant
mass calculation. 
The invariant mass window used to select $J/\psi$ candidates in the $\mu ^+ \mu ^-(e ^+ e ^-)$ channel is 
$-0.06~(-0.15)~{\rm GeV/}c^2~\le M_{\ell^+\ell^-} - m_{J/\psi} \le 0.036~{\rm GeV/}c^2$, where $m_{J/\psi}$ denotes the nominal $J/\psi$ mass~\cite{Hag}; these intervals are asymmetric in order to include part of the radiative tails. 
 Vertex- and mass-constrained kinematic fits are performed for selected $J/\psi$ candidates to improve the momentum resolution. 

Photons are identified as ECL energy clusters that are not associated with a charged track and  have a minimum energy of $0.060~{\rm GeV}$. 
We reject the 
photon candidate if the ratio of the energy in the array of the central $3\times 3$ ECL cells to that in the array of $5\times 5 $ cells is less than 0.87. 

To reconstruct the $\chi_{c1}$ state, we combine a $J/\psi$ candidate with momentum below 2.0~GeV/$c$ in the CM frame with a selected photon. 
To suppress  photons arising from $\pi^0 \to \gamma \gamma$, we veto photons that, when combined with another photon in the event, satisfy $0.110~{\rm GeV/}c^2 \le M_{\gamma\gamma}\le 0.150~{\rm GeV/}c^2$. 
The $\chi_{c1}$ candidates are selected by
requiring the mass difference ($\Delta M~=~M_{\ell ^+ \ell ^-\gamma} - M_{\ell^+\ell^-}$) to lie between $0.370~{\rm GeV/}c^2$ and $0.438~{\rm GeV/}c^2$. The $\Delta M$ distribution is shown in Fig. 2. A mass-constrained fit is applied to all selected $\chi _{c1}$ candidates in order to improve the momentum resolution. 

Charged pions and kaons are identified using energy loss measurements in the CDC, Cherenkov light yields in the ACC, and TOF information. The information from these detectors is combined to form $\pi$-$K$ likelihood ratio, $\mathcal R(\pi/K) = \mathcal L_{\pi}/(\mathcal L_{\pi} +\mathcal L_{K})$, where $\mathcal L_{\pi}(\mathcal L_{K})$ is the likelihood that a pion (kaon) would produce the observed detector response. 
Charged tracks with $\mathcal R(\pi/K) > 0.9$ are selected as charged pions, and tracks with $\mathcal R(\pi/K) \leq 0.4$ are selected as charged kaons. 
The efficiency for pion (kaon) identification is 75.1\% (86.1\%) and the probability of kaon (pion) misidentification is 4.6\% (10.5\%) with the above criteria. We determine the selection criteria by optimizing the figure of merit, $S/\sqrt(S+B)$, where $S (B)$ is the number of signal (background) events in the signal region, with an assumed branching fraction that is 5\% of that for $B^- \to \chi_{c1} K^-$~\cite{Hag}.


We reconstruct $B$-mesons by combining a $\chi_{c1}$ candidate with a charged pion or kaon. The energy difference, $\Delta E \equiv E_{B}^*- E_{\rm beam}^*$ and the beam-constrained mass $M_{\rm bc} \equiv \sqrt{E_{\rm beam}^{*2} - p_B^{*2}}$, are used to separate signal from background, where $E_{\rm beam}^{*}$ is the run dependent beam energy, and $E_{B}^*$ and $p_{B}^*$ are the reconstructed energy and momentum, respectively of the $B$-meson candidates in the CM frame. 
We accept candidates in the region 
5.27~${\rm GeV/}c^2 \leq M_{\rm bc} \leq$ 5.29 $~{\rm GeV}/c^{2}$ and
$|\Delta E| < 0.2~(0.15)~\rm GeV$ for the $B^-\to \chi _{c1} \pi^- (K^-)$ mode.
When an event contains more than one $B$-meson candidate passing the above requirements (this occurs in $\sim$ 2.5\% of the candidate events), the candidate with $M_{\rm bc}$ closest to the nominal $B^-$ mass~\cite{Hag} is selected.

We extract the signal yields by performing a binned maximum likelihood fit to 
the $\Delta E$ distribution of the selected candidates.
For the $B^- \to \chi_{c1}K^-$ mode, we fit with a sum of two Gaussians for signal and a second-order polynomial for background.
In the fit for the $B^-\to \chi_{c1}\pi^-$ mode, a background component exists due to misidentified 
$\chi_{c1}K^-$.
This background has a peak at $\Delta E \sim -0.07$ GeV 
and is modeled by a sum of two bifurcated Gaussians. 
A third-order polynomial is used for the sum of all other backgrounds.
We study backgrounds using a large sample of inclusive charmonium
Monte Carlo (MC) events \cite{evtgen}.  
Except for the misidentified $B^- \to \chi_{c1}K^-$ background,
no structure is observed in the $\Delta E$ distribution. However, the $M_{\rm bc}$ distribution has a peaking component. Therefore, we use the $\Delta E$
distributions for the signal extraction. The scatter plot of $\Delta E$ versus $M_{\rm bc}$ for $B^- \to \chi_{c1} \pi^-$ candidates is shown in Fig. 3, where the $M_{\rm bc}$ requirement is loosened to 5.2~{\rm GeV/c$^2$}.

All parameters of the fitting functions are floated in the fit of the $B^-\to \chi_{c1}K^-$ mode.
For the $B^-\to \chi_{c1}\pi^-$ mode, the signal shape is fixed to that 
obtained from the $B^- \to \chi_{c1}K^-$ mode.  The shape of misidentified $\chi_{c1}K^-$
background is initially determined from a MC sample, with a correction applied to account for the small difference between data and MC in
the $B^-\to \chi _{c1} K^-$ sample.
We obtain $1597\pm48$ and $55\pm10$ signal events for  
$B^-\to \chi _{c1} K^- $ and $B^- \to \chi _{c1} \pi^-$ modes, respectively.
The $\Delta E$ distributions are shown in Figs. 4 and 5, together with
the fit results.
The number of  of misidentified $B^- \to \chi_{c1}K^-$ events obtained from the fit to Fig. 5 is $61 \pm 14$. This is consistent with the expectation from the observed $B^-\to \chi _
{c1} K^-$ signal yield (Fig. 4) given the probability of misidentifying
a kaon as a pion.
The significance of the $B^-\to \chi _{c1} \pi^-$ signal is 6.3$\sigma$, 
where the significance is defined as 
$\sqrt{-2\ln(\mathcal L_0/\mathcal L_{\rm max})}$ and 
$\mathcal L_{\rm max}$ ($\mathcal L_0$) denotes the likelihood value
at the maximum (with the signal yield fixed at zero). 
We include the effect of systematic error in this calculation by
subtracting a quadratic sum of the variations of the significance in
smaller direction when each fixed parameter in the fit is changed by 
$\pm 1\sigma$.
\begin{figure}
\begin{tabular}{c}
\resizebox{8cm}{6cm}{\includegraphics{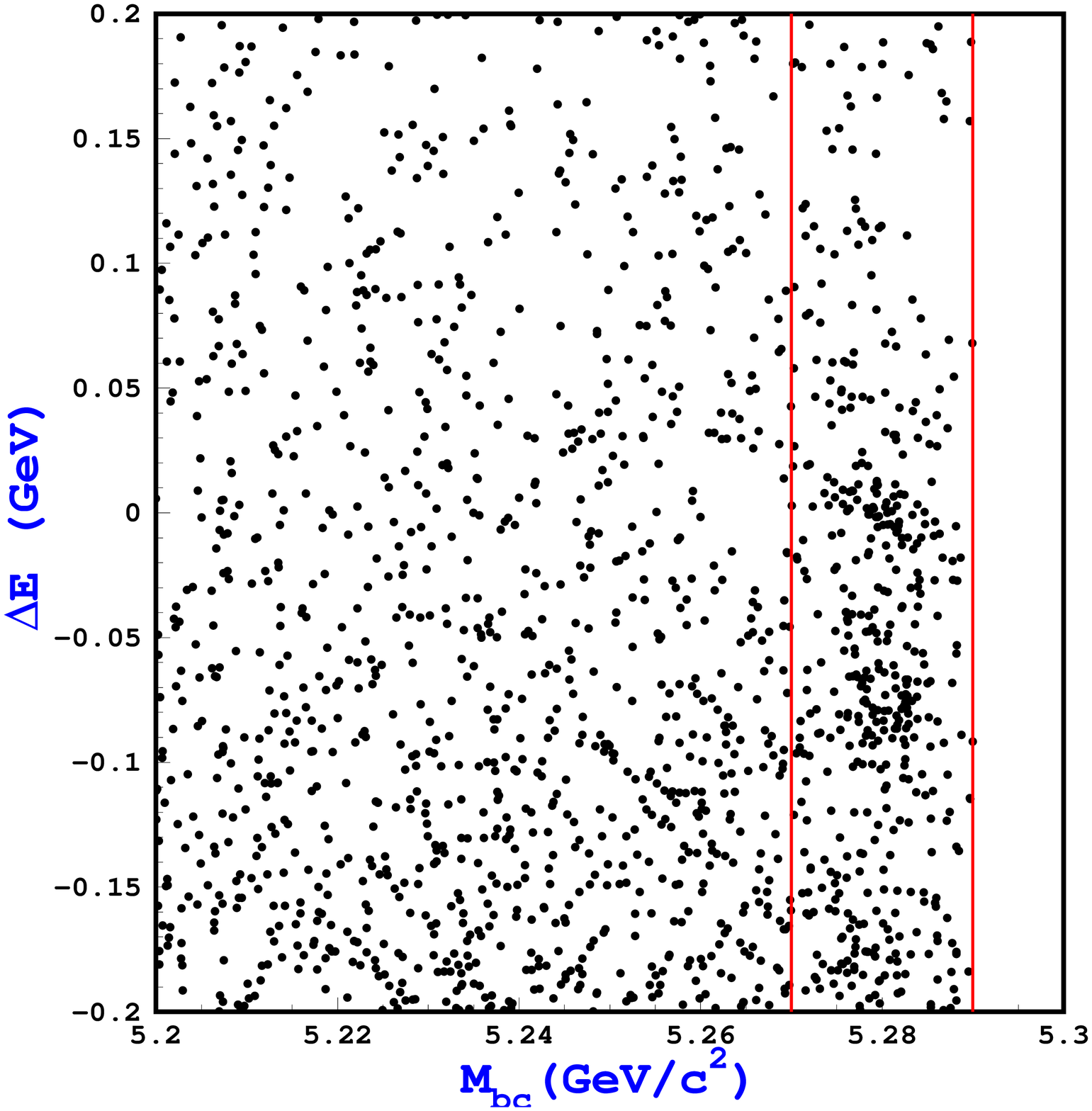}}
\end{tabular}
\caption{The scatter plot of $\Delta E$ versus $M_{\rm bc}$ for $B^{\pm} \to \chi_{c1} \pi^{\pm}$ candidates, where the two vertical lines indicate the $M_{\rm bc}$ region used for signal extraction.}
\end{figure}

\begin{figure}
\begin{tabular}{c}
\resizebox{8cm}{6cm}{\includegraphics{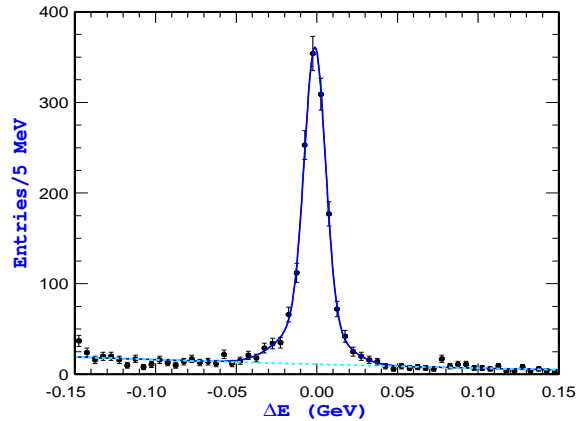}}
\end{tabular}
\caption{ The $\Delta E$ distribution for the $B^{\pm} \to \chi_{c1} K^{\pm}$ decay mode. The solid and dashed curves show the total fit and the polynomial background component of the fit, respectively.}
\end{figure}
\begin{figure}
\begin{tabular}{c}
\resizebox{8cm}{6cm}{\includegraphics{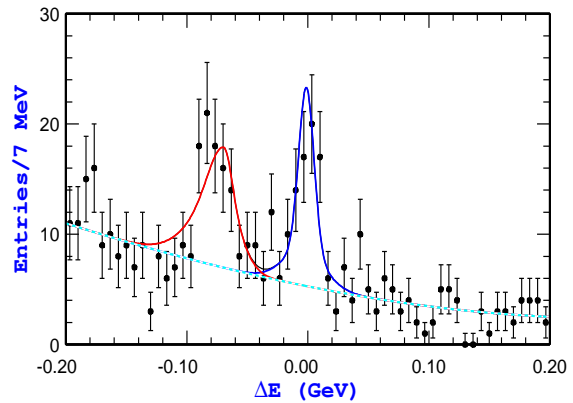}}
\end{tabular}
\caption{The $\Delta E$ distribution for the $B^{\pm} \to \chi_{c1} \pi^{\pm}$ decay mode. The signal peak is seen around zero. The peak at $-0.07~{\rm GeV}$ is from $B^{\pm} \to \chi_{c1} K^{\pm}$ decay. The solid and dashed curves show the total fit and the polynomial background component of the fit, respectively.}
\end{figure}

The branching fraction for the $B^-\to {\chi}_{c1}\pi^- $ decay mode
 is calculated by dividing the observed signal yield by the
 reconstruction efficiency, the number of $B{\overline B}$ events in the
 data sample, and the daughter branching fractions. 
We determine the reconstruction efficiency (17.3\%) from signal MC events, where the correction for difference between data and MC has been applied for the pion identification requirement (0.90$\pm$0.01) and the $\Delta M$ requirement (0.97$\pm$0.03). The correction factor for the $\Delta M$ requirement is determined from the $B^-\to \chi_{c1} K^-$ sample and is estimated by taking the ratio of yields from data and MC for tight ($0.370~{\rm GeV/}c^2 <\Delta M < 0.438~{\rm GeV/}c^2$)  and loose ($0.3~{\rm GeV/}c^2 <\Delta M < 0.5~{\rm GeV/}c^2$) $\Delta M$ windows. We use the daughter branching fractions published in Particle Data Book 2004~\cite{Hag}. 
Equal production of neutral and charged $B$-meson pairs in
 $\Upsilon ({\rm 4S})$ decay is assumed. 
The resulting branching fraction is  
\begin{equation}
  \mathcal {B}( B^{\pm} \to \chi_{c1} \pi^{\pm}) 
     = (2.2\pm 0.4\pm 0.3)\times 10^{-5},
\end{equation}
where the first error is statistical and the second is systematic. 
We obtain the branching fraction for the $B^-\to \chi _{c1} K^-$ decay mode by similar procedures. The result, $(51.4 \pm 1.5) \times 10^{-5}$, (error is statistical only) 
is consistent with the previous measurements~\cite{bellechic, babar}.
The ratio of branching fractions is
\begin{equation}
\frac{\mathcal{B}(B^- \to \chi_{c1} \pi^-)}{\mathcal{B}(B^- \to \chi_{c1} K^-)}
      = (4.3\pm 0.8\pm 0.3)\%,
\end{equation}
which is consistent with expectations from the factorization model~\cite{five}.

The systematic uncertainties are summarized in Table~\ref{tab:table2}.
Since the shape used for the signal is fixed in the fit to Fig. 5, we
repeat the fit, varying each fixed shape parameter by the $\pm 1\sigma$
uncertainty in our determination of it from external samples (Fig. 4 and
MC).  The systematic uncertainty on the signal yield is then calculated
by taking the quadratic sum of the deviations in the signal yield from
the nominal value.
We checked for possible bias in the fitting using a MC sample; no significant bias was found. The systematic uncertainty assigned to the yield is 5.9$\%$.
The uncertainty on the tracking efficiency is estimated to be 1.0$\%$ per track, while that due to lepton identification is $2.0\%$ per lepton,
and $1.0\%$ ($1.3\%$) per pion (kaon) identification (PID). 
We assign an uncertainty of $2.0\%$ for the $\gamma$ detection efficiency. 
The systematic uncertainty due to ${\chi}_{c1}\to \gamma J/\psi$ and $J/\psi\to \ell^+ \ell^-$ branching fractions is 10.6$\%$.
The total systematic error is the sum of all the above uncertainties in quadrature.

Many of the systematic errors cancel for the ratio of branching fractions;
contributions come from only the uncertainty in the $B^-\to \chi _{c1} \pi^-$ 
yield, PID (1.0\% for $B^-\to \chi _{c1} K^-$), and MC statistics (1.0\% for $B^-\to \chi _{c1} K^-$).

\begin{table}
\caption{\label{tab:table2} Summary of systematic errors on
   branching fraction.   
}
\begin{ruledtabular}
\begin{tabular}{lc}                      
 Source & Uncertainty (\%)
\\\hline
Uncertainty in yield & 5.9   \\         
Tracking error & 3.0         \\         
Lepton Identification & 4.0              \\ 
PID (pion)& 1.0     \\          
$\gamma$ detection &2.0      \\          
$\Delta M$ requirement & 3.0 \\          
MC statistics & 0.9          \\          
$\rm {N}_{B\bar B}$ & 1.2    \\          
Daughter branching fractions & 10.6      
\\\hline
Total & 13.7                 \\          
\end{tabular} 
\end{ruledtabular}
\end{table}

The $CP$-violating charge asymmetry $\mathcal {A}_i$ is defined as \\
\begin{equation}
\mathcal{A}_i = \frac {N_{i}^{-} - N_{i}^{+}}{N_{i}^{-} + N_{i}^{+}}; ~~i = \pi, K.
\end{equation}
Here, $N_{i}^-$ and $N_{i}^+$ are the signal yields for negative and positive $B$-meson decays and are measured separately by using the method described above. 
For the $B^-\to \chi _{c1} \pi^-$ mode, the polynomial background shape is fixed to that obtained for the branching fraction measurement.

The measured charge asymmetries for 
the $B^{\pm}\to \chi _{c1} \pi^{\pm} (K^{\pm})$ decay modes are listed in 
Table~\ref{tab:table3}. 
No significant asymmetries are seen in either decay modes.
 The systematic errors on $\mathcal {A}_{\pi}$ ($\mathcal {A}_{K}$) include:
uncertainty in yield extraction, 0.007; possible difference between $B^-$ and $B^+$ signal shape parameters,
0.002 (0.001); possible charge asymmetry in pion (kaon) identification efficiency 0.014 (0.011); and possible detector bias 0.016, which is estimated from the charge asymmetry of the $B^{\pm}\to J/\psi K^{\pm}$ decay sample without a PID requirement.
\begin{table}
\caption{\label{tab:table3} Summary of charge asymmetries. 
 First (second) error is statistical (systematic).}
\begin{ruledtabular}
\begin{tabular}{cccc}
Mode & Yield($-$)& Yield(+) & $\mathcal A$\\\hline
$B^\pm\to \chi _{c1} \pi^\pm$& $29\pm7$& $25\pm7$&$~~0.07\pm0.18\pm0.02$\\
$B^\pm\to \chi _{c1} K^\pm$& $792\pm31$ & $807\pm31$ & $-0.01\pm0.03\pm0.02$\\
\end{tabular}
\end{ruledtabular}
\end{table}

In summary, we report the first observation of $B^- \to \chi_{c1}\pi^-$ decay with $386\times10^6$~$B\overline B$ events.~The observed signal yield is $55\pm10$ with a significance of 6.3$\sigma$ including systematic uncertainty.~The measured branching fraction is $\mathcal {B}( B^{\pm} \to \chi_{c1} \pi^{\pm}) = (2.2\pm 0.4\pm 0.3)\times 10^{-5}$.
The ratio ${\mathcal{B}(B^- \to \chi_{c1} \pi^-)}/{\mathcal{B}(B^- \to \chi_{c1} K^-)} = (4.3\pm 0.8\pm 0.3)\% $, which is consistent with the Standard Model prediction.
While the accuracy of $\mathcal {A}_{K}$ is improved from the previous 
measurement~\cite{babar}, no significant $CP$-violating charge asymmetries are observed in either $B^{\pm}\to \chi _{c1} \pi^{\pm}$ or
$B^{\pm}\to \chi _{c1} K^{\pm}$ decay modes.

We thank the KEKB group for the excellent operation of the
accelerator, the KEK cryogenics group for the efficient
operation of the solenoid, and the KEK computer group and
the National Institute of Informatics for valuable computing
and Super-SINET network support. We acknowledge support from
the Ministry of Education, Culture, Sports, Science, and
Technology of Japan and the Japan Society for the Promotion
of Science; the Australian Research Council and the
Australian Department of Education, Science and Training;
the National Science Foundation of China and the Knowledge
Innovation Program of the Chinese Academy of Sciencies under
contract No.~10575109 and IHEP-U-503; the Department of
Science and Technology of India; 
the BK21 program of the Ministry of Education of Korea, 
the CHEP SRC program and Basic Research program 
(grant No.~R01-2005-000-10089-0) of the Korea Science and
Engineering Foundation, and the Pure Basic Research Group 
program of the Korea Research Foundation; 
the Polish State Committee for Scientific Research; 
the Ministry of Science and Technology of the Russian
Federation; the Slovenian Research Agency;  the Swiss
National Science Foundation; the National Science Council
and the Ministry of Education of Taiwan; and the U.S.\
Department of Energy.


\begin{thebibliography}{99}
\bibitem{foot}
Charge-conjugate modes are included throughout this paper unless stated otherwise.
\end{thebibliography}

\end{document}